\documentclass[aps,prb,twocolumn,showpacs,preprintnumbers,superscriptaddress,amsmath,amssymb,10pt]{revtex4-1}

\usepackage[T1]{fontenc}
\usepackage{mathcomp}
 
\usepackage{rotating}

\usepackage[justification=raggedright,singlelinecheck=false]{caption}

\begin{document}

\title{Photoluminescence of freestanding single- and few-layer MoS$_2$}

\author{Nils Scheuschner}\email{nils.scheuschner@tu-berlin.de}
\affiliation{Institut für Festkörperphysik, Technische Universität Berlin, Hardenbergstr. 36, 10623 Berlin, Germany}
\author{Oliver Ochedowski}
\affiliation{Fakultät für Physik und CeNIDE, Universität Duisburg-Essen, 47048 Duisburg, Germany}
\author{Anne-Marie Kaulitz}
\affiliation{Institut für Festkörperphysik, Technische Universität Berlin, Hardenbergstr. 36, 10623 Berlin, Germany}
\author{Roland Gillen}
\affiliation{Institut für Festkörperphysik, Technische Universität Berlin, Hardenbergstr. 36, 10623 Berlin, Germany}
\author{Marika Schleberger}
\affiliation{Fakultät für Physik und CeNIDE, Universität Duisburg-Essen, 47048 Duisburg, Germany}
\author{Janina Maultzsch}
\affiliation{Institut für Festkörperphysik, Technische Universität Berlin, Hardenbergstr. 36, 10623 Berlin, Germany}

\date{\today}

\begin{abstract}

We present a photoluminescence study of freestanding and Si/SiO$_2$ supported single- and few-layer MoS$_2$. The single-layer exciton peak (\textit{A}) is only observed in freestanding MoS$_2$. The photoluminescence of supported single-layer MoS$_2$ is instead originating from the \textit{A}$^-$ (trion) peak as the MoS$_2$ is \textit{n}-type doped from the substrate.  In bilayer MoS$_2$, the van der Waals interaction with the substrate is decreasing the indirect band gap energy by up to $\approx$ 80 meV. Furthermore, the photoluminescence spectra of suspended MoS$_2$ can be influenced by interference effects.

\end{abstract}

\pacs{
78.55.Hx, 
73.21.Ac, 
78.20.Ci, 
78.30.Hv 
}

\maketitle

\section{Introduction}

Two-dimensional crystals of molybdenum disulfide (MoS$_2$) show a great potential for novel nanoelectronic devices.\cite{Radisavljevic2011,Lembke2012} Bulk MoS$_2$ is an indirect semiconductor with a band gap of 1.29 eV.\cite{Gmelin} In absorption it shows two excitonic transitions \textit{A} ($\approx$ 1.8 eV) and \textit{B} ($\approx$ 2.0 eV) from the \textit{K} point of the Brillouin zone \cite{Evans1965}. As bulk MoS$_2$ is an indirect semiconductor, it shows virtually no photoluminescence. Reducing the thickness of the MoS$_2$ to a few layers leads to a strong increase in photoluminescence yield\cite{Mak2010,Splendiani2010}. The emerging photoluminescence is attributed to the increase of the indirect band gap due to confinement effects, which leads to a transition to a direct semiconductor for single-layer MoS$_2$\cite{Mak2010,Splendiani2010,Cheiwchanchamnangij2012}. This change in the electronic band structure was also directly measured in angle-resolved photoemission spectroscopy.\cite{Jin2013} 

Figure \ref{Sample}(a) shows photoluminescence spectra of single- and bilayer MoS$_2$ on Si/SiO$_2$ substrate. For more than one layer, an additional low-energy feature \textit{I} is observed, which is attributed to the indirect gap.\cite{Mak2010} As predicted by band structure calculations, the peak position of \textit{I} strongly depends on the layer number, see Fig. \ref{Sample}(b). Mak \textit{et al.} \cite{Mak2010} reported strong photoluminescence of freestanding single-layer MoS$_2$ with a peak position of 1.90 eV. They found for the freestanding single-layer MoS$_2$ an increase in photoluminescence intensity by two orders of magnitude compared to the bilayer. For MoS$_2$ on Si/SiO$_2$ substrates, on the other hand, an increase of only approximately 40\% was reported; the photoluminescence peak position was determined to 1.83 eV.\cite{Splendiani2010} These large differences in quantum yield and transition energy raise the question how the substrate influences the optical properties of single- and few-layer MoS$_2$. 

In this work we present the direct comparison of the photoluminescence of freestanding and supported (Si/SiO$_2$ substrates) single- and few-layer MoS$_2$. For few-layer MoS$_2$, we show that, besides the confinement effects leading to the indirect-direct transition \cite{Mak2010,Splendiani2010}, the van der Waals interaction with the substrate has a strong influence on the electronic structure at the $\Gamma$ point. This results in a blue shift of the indirect transition of up to 80 meV in bilayer MoS$_2$. 
For single-layer MoS$_2$ on Si/SiO$_2$ substrates, the photoluminescence peak is at $\approx$ 1.82 eV; for freestanding samples the peak blueshifts by $\approx$ 65 meV and its intensity increases by up to one order of magnitude. We attribute this to the simultaneous observation of the \textit{A} exciton and the \textit{A}$^-$ peak (attributed to negatively charged trions\cite{Mak2012c}) in freestanding single-layer MoS$_2$. As the exciton emission is suppressed depending on the charge carrier concentration, we conclude that the single-layer MoS$_2$ on Si/SiO$_2$ substrates is effectively \textit{n}-type doped by the substrate. This finding implies that in most cases where photoluminescence of single-layer MoS$_2$ in the energy range of $\approx$ 1.82 eV has been reported, the MoS$_2$ was \textit{n}-type doped and the observed emission was originating primarily from the \textit{A}$^-$ peak (trion) and not from the exciton (\textit{A}).

\section{Samples and Methods}

\begin{figure}
\begin{minipage}{0.5\textwidth}
\includegraphics*[width=1.0\textwidth]{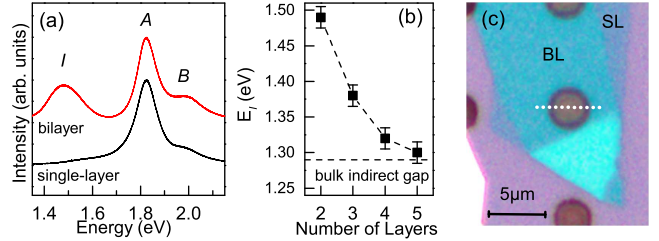}
\end{minipage}
\caption{\label{Sample} (a) Photoluminescence spectra of single-layer and bilayer of MoS$_2$ on Si/SiO$_2$ substrate excited with 2.33 eV. The transitions \textit{I}, \textit{A} and \textit{B} are indicated; spectra are vertically offset. (b) Indirect transition energy of few-layer MoS$_2$ (E$_I$) on Si/SiO$_2$ substrates.
(c) Optical image of a single- (SL) and bilayer (BL) sample on a Si/SiO2 substrate with holes; the dotted line indicates the linescan shown in Fig. \ref{BL_Line}.}

\end{figure}

We prepared freestanding MoS$_2$ layers via mechanical exfoliation of natural MoS$_2$ (SPI supplies) on Si/SiO$_2$ substrates. The thickness of the SiO$_2$ is 100 nm, which leads to an increased optical contrast of the atomically thin MoS$_2$ flakes \cite{Castellanos-Gomez2010}. A regular pattern of holes with diameters >3 $\mu$m was etched into the wafers. Via optical microscopy we identified single- and few-layer MoS$_2$ flakes covering the Si/SiO$_2$ substrate as well as holes, see Fig. \ref{Sample}(c). The layer number was verified by Raman and photoluminescence spectroscopy\cite{Lee2010c}.
Micro-Raman and photoluminescence microscopy were performed with a Horiba Labram HR spectrometer with 2.33 eV excitation energy. The photoluminescence spectra were corrected for the relative spectral sensitivity of the system with a National Institute of Standards and Technology (NIST) traceable reference white light source. The laser intensity was chosen below 0.5 mW. The laser spot size was $\approx$  0.5 $\mu$m. All measurements were performed at room temperature.

\section{Results and Discussion}

\begin{figure}
\begin{minipage}{0.5\textwidth}
\includegraphics*[width=1.0\textwidth]{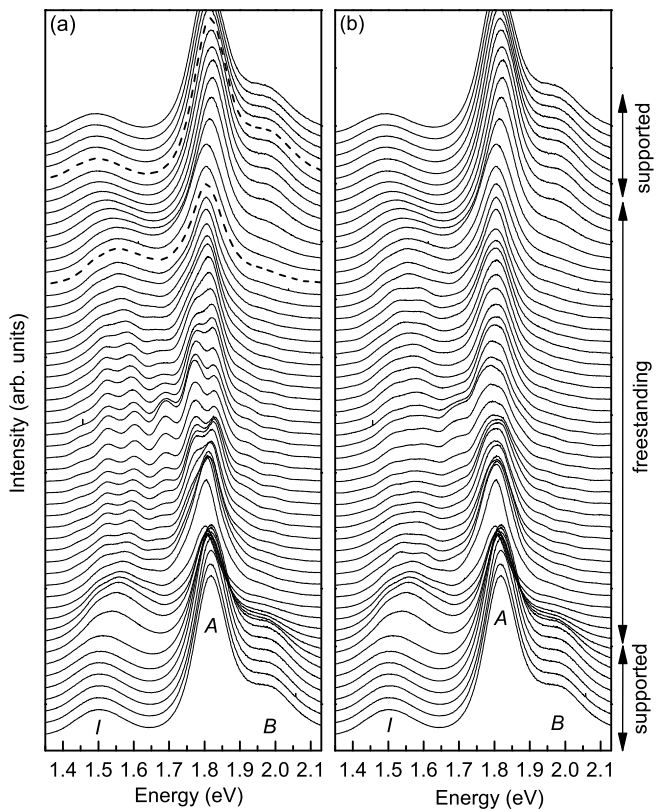}
\end{minipage}
\caption{\label{BL_Line} Photoluminescence linescan of a bilayer MoS$_2$ sample across supported and freestanding areas, following the dotted line in Fig. \ref{Sample}(c). (a) Unprocessed spectra. The two dashed lines indicate the spectra shown in Fig. \ref{BL_I}(a). (b) Demodulated spectra.}
\end{figure}

Figure \ref{BL_Line}(a) shows the evolution of the photoluminescence spectra of bilayer MoS$_2$ in a linescan from the substrate across a freestanding area, as indicated by the dotted line in Fig. \ref{Sample}(c). At first sight, the most apparent change is that the spectra from the freestanding area show a modulation of the signal with a periodicity $\Delta E$ of $\approx$ 80 meV. Considering the geometry of the system, we suggest the modulation to be caused by interference of the directly emitted light with the light reflected from the bottom of the hole. In this case, $\Delta E$ can be calculated by $\Delta E = \frac{hc}{2L}$, where $L$ is depth of the hole, $h$ is Planck's constant and $c$ is the velocity of light. By optical microscopy of the cross section of one of our substrates, we determine $L$ $\approx$ 8 $\mu$m, which leads to a predicted value for $\Delta E$ of 77 meV. This is in excellent agreement with the observed value. Therefore, to correct our data for further analysis, we introduce an empirical modulation function $f_{\text{mod}}$:
\begin{equation}
f_{\text{mod}}(E)=1+A_{\text{mod}} \cdot\cos\left(E\cdot\frac{2\pi}{\Delta E}+\varphi_{\text{mod}}\right)
\end{equation}
$A_{\text{mod}}$ describes the amplitude and $\varphi_{\text{mod}}$ the phase of the modulation. On the substrate, the photoluminescence spectrum of bilayer MoS$_2$ can be fitted with two Lorentzians for the $B$ and $I$ transitions, one Gaussian for the $A$ transition, and a constant background term, see dashed line in Fig. \ref{BL_I}(a). We multiply the modulation function $f_{\text{mod}}$ to these three peaks to fit the freestanding bilayer spectra. With the results of  $\Delta E$, $A_{\text{mod}}$ and $\varphi_{\text{mod}}$ from the fits, we then demodulate the photoluminescence spectra as shown in Fig. \ref{BL_Line}(b). 
The amplitude of the modulation varies strongly between the different holes on the substrate. We attribute this effect to varying roughness and tilting of the bottoms of the holes.

\subsection{Freestanding few-layer MoS$_2$}

\begin{figure}
\begin{minipage}{0.5\textwidth}
\includegraphics*[width=1.0\textwidth]{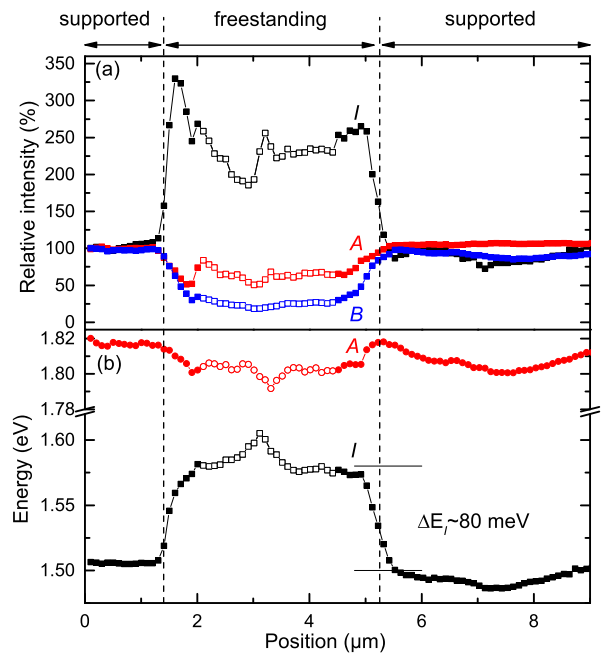}
\end{minipage}
\caption{\label{BL_extrahiert} (a) Indirect and A exciton transition energy extracted from the linescan for bilayer MoS$_2$. (b) \textit{A}, \textit{B} and \textit{I} photoluminescence intensities normalized to the substrate values. Data points with filled (hollow) symbols were determined from the fits of the unprocessed (demodulated) spectra.}
\end{figure}

Figure \ref{BL_extrahiert}(a) shows the relative photoluminescence intensities of the \textit{A}, \textit{B} and \textit{I} transitions normalized to the substrate values for the linescan over the bilayer sample. We define the intensity of the transitions as the area under the Lorentzian (\textit{B} and \textit{I}) or Gaussian (\textit{A}) curve. The \textit{A} exciton intensity for the freestanding parts is reduced to 60-70\% of the intensity in the supported region; we also observe an analogous reduction of the Raman intensity. 
Such a decrease is expected compared to the supported part, where multiple reflections of the exciting and the emitted light at the Si/SiO$_2$ interface lead to an enhancement.\cite{Li2012d} The intensity of the \textit{B} transition is at least reduced to 20\%, as it can already be seen directly in Fig. \ref{BL_Line}(a). In contrast, the intensity in the freestanding part of the \textit{I} transition is increased to 200-320\%. From our measurements the reason for this intensity change remains unclear.

\begin{figure}
\begin{minipage}{0.5\textwidth}
\includegraphics*[width=1.0\textwidth]{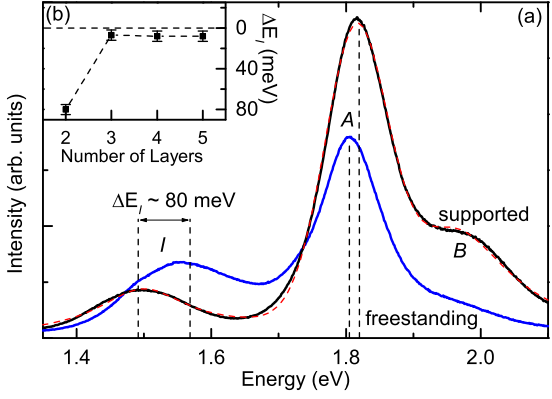}
\end{minipage}
\caption{\label{BL_I} (a) Photoluminescence spectra of freestanding and supported bilayer MoS$_2$. The spectrum of the freestanding bilayer is from the region near the edge of the hole, where only very small modulation effects are observed; The dashed line shows the fitting function for the supported bilayer MoS$_2$ spectrum. (b) Difference of the indirect transition energy ($\Delta$E$_I$) of few-layer MoS$_2$ between supported and freestanding areas as a function of the layer number.}
\end{figure}

Figure \ref{BL_extrahiert}(b) shows the energies of the \textit{A} and \textit{I} peaks. In freestanding bilayer MoS$_2$, we observe an increase in the energy of the \textit{I} transition by $\approx$ 80 meV, and a decrease of the \textit{A}-transition energy by $\approx$ 15 meV, compared to the surrounding supported areas. Those shifts are already evident in the spectra of the freestanding area close to the supported area, which show no modulation effects, see Figs. \ref{BL_extrahiert}(a) and  \ref{BL_I}(a).
We will discuss in the following three scenarios as potential source of these shifts: $(i)$ the dielectric environment\cite{Mao2013,Plechinger2011}, $(ii)$ strain\cite{Conley2013a,He2013,Shi2013,Zhu2013} and $(iii)$ direct changes in the electronic band structure due to interaction with the substrate.

$(i)$ It is well known that the dielectric environment can change the exciton binding energy by screening of the Coulomb interaction and modifying electron-electron interactions. However, we conclude from our measurements that the \textit{A} and \textit{I} peaks are similarly affected by changes in the dielectric environment: On the substrate the individual peak positions show standard deviations of $\approx$ 7 meV from the center position, while their difference shows only a standard deviation of $\approx$ 2 meV, \textit{i.e.}, both peaks shift in the same direction. Assuming that these shifts are due to local variations in the dielectric environment as well (inhomogeneities in the substrate, adsorbates on the MoS$_2$ flakes), the change of the dielectric environment cannot explain the opposite shifts of the \textit{A} and \textit{I} peaks when going from the substrate to the freestanding part.

\begin{figure}
\begin{minipage}{0.5\textwidth}
\includegraphics*[width=1.0\textwidth]{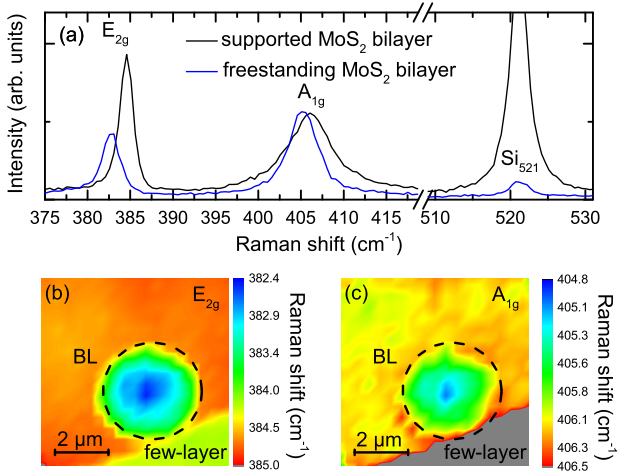}
\end{minipage}
\caption{\label{RamanBL} (a) Raman spectra of freestanding and supported bilayer MoS$_2$. (b) and (c) Raman mappings showing the E$_{2g}$ (a) and A$_{1g}$ (b) Raman mode positions. The dashed lines indicate the position of the hole.}
\end{figure}

$(ii)$ Raman spectroscopy shows a decrease of the E$_{2g}$ phonon frequency by 2 cm$^{-1}$ in the freestanding areas relative to the supported area, see Fig. \ref{RamanBL}. This shift might be due to biaxial strain as the MoS$_2$ might sag in the hole area. 
To assess the effect of such slight sagging on the phonon frequencies, we simulated their evolution under applied biaxial tensile strain $\epsilon$ using density functional theory (DFT)\footnote{The phonon frequencies at the $\Gamma$-point and the electronic band structures were calculated in the frame of density functional (perturbation) theory on the level of the Perdew-Burke-Ernzerhof (PBE) exchange-correlation functional as implemented into the Quantum Espresso suite. We treated the Mo(3s,3p,3d,42) and the S(3s,3p) states as valence electrons using projector augmented waves (PAW) with a cutoff of 70 Ry. All reciprocal space integrations were performed by a discrete k-point sampling of 12x12x1 k-points in the Brillouin zone. For each hydrostatic strain $\epsilon$, we scaled the calculated in-plane lattice constant of a=3.19\AA\space of the unstrained MoS$_2$ bilayer sheet to a'=(1+$\epsilon$)a and fully optimized the atomic positions until the residual forces between atoms was smaller than 0.01 eV/\AA. Interactions of the sheet with residual periodic images due to the 3D boundary conditions were minimized by maintaining a vacuum layer between bilayer planes of at least 25\AA.}. 
For the purpose of this work, we considered a representative range of $\epsilon$ = 0$-$1\%, where we defined $\epsilon$ as the relative length change of the lattice vectors, \textit{i.e.}, $\epsilon$=$\frac{\Delta a}{a}$, due to in-plane hydrostatic stress $\sigma\equiv\sigma_{xx}=\sigma_{yy}$. As expected all modes experience a red shift for tensile strain, with a shift rate of $\approx-$2.2 cm$^{-1}/\%$ for the A$_{1g}$ mode and $\approx-$5.2 cm$^{-1}/\%$ for the E$_{2g}$ mode, respectively. Assuming that the Raman E$_{2g}$ mode shows almost no dependence on doping\cite{Chakraborty2012,Frey1999}, the observed downshift mode would correspond to biaxial tensile strain of $\epsilon\approx$ 0.4\%.

\begin{figure}
\begin{minipage}{0.5\textwidth}
\includegraphics*[width=1.0\textwidth]{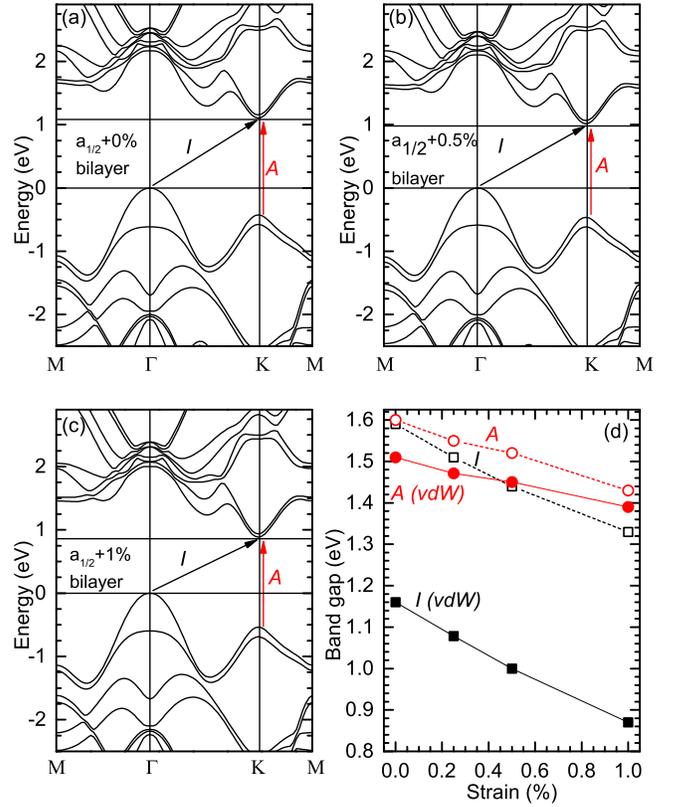}
\end{minipage}
\caption{\label{bands} Evolution of calculated electronic band structures for bilayer MoS$_2$ subject to biaxial tensile strain of (a) 0\%, (b) 0.5\% and (c) 1\% as calculated using van-der-Waals corrected DFT-PBE calculations. (d) Comparison of the trends of the corresponding indirect and direct band gaps for calculations with (full symbols) and without (open symbols) van-der-Waals (vdW) corrections.
The arrows denote the indirect and direct fundamental band gaps of the material.
}
\end{figure}

Tensile uniaxial strain causes redshifts of the \textit{A} and \textit{I} transitions. \cite{Conley2013a,He2013} Quasiparticle band structure calculations of monolayer MoS$_2$ under biaxial strain predict similar results.\cite{Shi2013} For few layer MoS$_2$, the influence of the van der Waals interaction between the layers, which leads to the direct-indirect gap transition, has to be taken into account. It appears feasible that this additional term could lead to opposite shifts of the \textit{A} and \textit{I} transitions of bilayer MoS$_2$ under biaxial strain. We thus calculated the corresponding electronic band structures of the biaxially strained MoS$_2$ bilayer sheets, with and without van der Waals interaction, see Fig.~\ref{bands}. The van der Waals interaction is primarily influencing the electronic states at the $\Gamma$ point, leading to a redshift of the \textit{I} transition by 0.43 eV for unstrained bilayer MoS$_2$, compared to calculations without van der Waals interaction. This redshift increases slightly for 1\% biaxial tensile strain to 0.46 eV. As the exciton binding energy is only marginally affected by small strain\cite{Feng2012}, we can assume that shifts of the photoluminescence peaks almost purely arise from strain-induced changes of the fundamental band gaps. With van der Waals interactions included, both the indirect and the direct fundamental band gaps decrease linearly in the range of 0$-$1\% strain with a rate of $-0.12$ eV/\% for the direct transition and a larger rate of $-0.29$ eV/\% for the indirect transition. Again, as both transitions shift to lower energies, tensile biaxial strain cannot explain the observed peak shifts in opposite directions.  

$(iii)$ As strain and the dielectric environment cannot explain the shifts of the \textit{A} and \textit{I} peaks, we attribute the observed blue shift of the \textit{I} peak to the sensitivity of the electronic $\Gamma$ point states to interactions with the environment. This sensitivity results from the spatial localization of the electronic states toward the outer regions of the MoS$_2$ layer.\cite{Splendiani2010}  In multilayer MoS$_2$, it leads to the direct-indirect band gap transition because of the layer-layer interaction\cite{Cheiwchanchamnangij2012,Han2011}, as supported by the van der Waals calculations in Fig. \ref{bands}(d). We conclude from our measurements that the interaction of the supported bilayer MoS$_2$ with adjacent amorphous SiO$_2$, similar to the interaction with another MoS2 layer, leads to a redshift of the \textit{I} transition. The influence of the substrate decreases strongly with the layer number: for more than two layers the shift of the \textit{I} peak is less than 10 meV, see Fig. \ref{BL_I}(b). 

In analogous measurements with two- to five-layer MoS$_2$, we did not find a systematic layer number dependence for the \textit{A} peak position or the shift of the \textit{A} peak position between freestanding and supported areas. A possible explanation for this observation might be the influence of different concentrations of impurities on the photoluminescence. This seems plausible as we use MoS$_2$ from natural sources. For different few-layer MoS$_2$ samples on Si/SiO$_2$ substrates, we found variations in the energy of the \textit{I} peak of up to 30 meV between different flakes, indicating that the strength of the interaction with the substrate is also varying.

Our findings have implications on the physics of multilayer heterostructures of two-dimensional crystals\cite{Geim2013}: these results show that in those structures the interaction between the layers can alter the electronic states of each layer such that they cannot be treated as isolated layers.

\subsection{Freestanding single-layer MoS$_2$}

\begin{figure}
\begin{minipage}{0.5\textwidth}
\includegraphics*[width=1.0\textwidth]{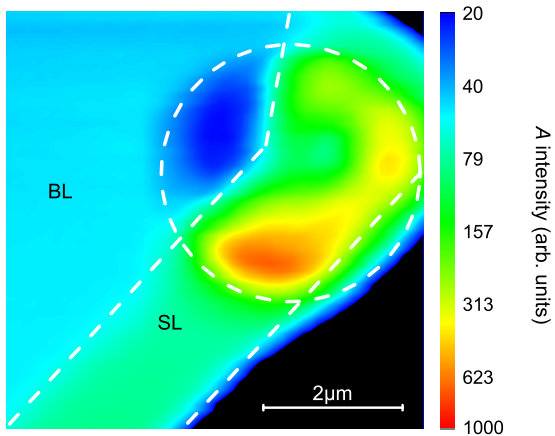}
\end{minipage}
\caption{\label{SL_map} Photoluminescence intensity map of the \textit{A} transition of a sample with freestanding single- and bilayer MoS$_2$. The dashed lines indicate the position of the hole and the border between single- (SL) and bilayer (BL) as observed in optical contrast.}
\end{figure}

Figure \ref{SL_map} shows a photoluminescence intensity map in the energy region of the \textit{A} transition of a sample with freestanding bilayer and single-layer MoS$_2$. When going from the supported to the freestanding region, the intensity of the \textit{A} transition decreases for the bilayer, whereas the intensity for the single-layer increases by up to one order of magnitude. This increase cannot be caused simply by an increase of the optical absorption cross section, as we observe a small decrease of the Raman intensity, which also contains the optical absorption cross section. In this particular sample, we observed only very weak modulation of the photoluminescence signal. 

\begin{figure}
\begin{minipage}{0.5\textwidth}
\includegraphics*[width=1.0\textwidth]{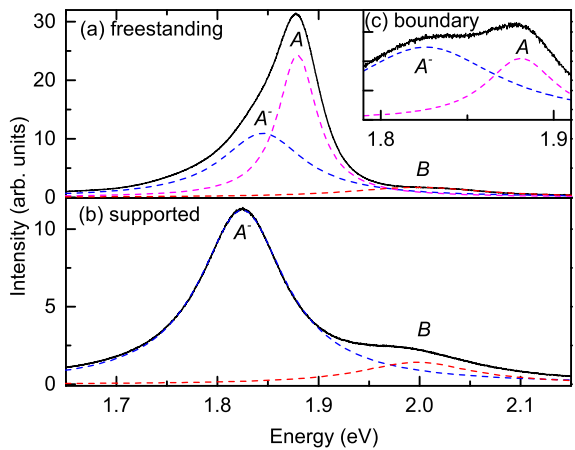}
\end{minipage}
\caption{\label{SL_Spektren} Photoluminescence of single-layer MoS$_2$ (solid lines). Dotted lines show fit functions of the \textit{A}, \textit{B} and \textit{A}$^-$-peaks. (a) Freestanding and (b) on Si/SiO$_2$ substrate and (c) from the boundary region between supported and freestanding areas, showing an overlap of both types of spectra.}
\end{figure}

Figure \ref{SL_Spektren} shows photoluminescence spectra from the freestanding (a) and supported (b) areas as well as from the boundary between the supported and freestanding areas (c). The maximum of the photoluminescence emission shows a blue shift of $\approx$ 65 meV in the freestanding single-layer compared to the supported one; the emission peak becomes asymmetric. At the boundary between supported and freestanding areas, the emission shows double-peak structures, see Fig. \ref{SL_Spektren}(c). We attribute this to the simultaneous observation of the \textit{A} peak and the \textit{A}$^-$ peak. The \textit{A}$^-$ peak was assigned by Mak \textit{et al.}\cite{Mak2012c} to negatively charged trions. While the emission intensity of the trions (\textit{A}$^-$) showed no dependence on the charge carrier concentration, the exciton (\textit{A}) emission was strongly reduced for \textit{n}-type doped MoS$_2$.\cite{Mak2012c}  For graphene it is well known that the charge carrier concentration can be influenced by the substrate.\cite{Stampfer2012,Bukowska2011} The observed changes in the photoluminescence of freestanding single-layer MoS$_2$ can thus be understood when assuming \textit{n}-type doping of the MoS$_2$ by charge transfer effects from the substrate. Therefore, the emission of the exciton (\textit{A}) is suppressed on the substrate; instead, the observed photoluminescence of single-layer MoS$_2$ on Si/SiO$_2$ substrate is primarily from the trion (\textit{A}$^-$). In the freestanding areas, the MoS$_2$ layer is less doped, and the emission of the exciton becomes dominant. This is seen in the increase in intensity and the observed blue shift in Fig. \ref{SL_Spektren}(a). From the photoluminescence map we determine the energy of the excitonic \textit{A} transition of freestanding single-layer MoS$_2$ to $1.886\pm 0.008$ eV. As the exciton is completely quenched on the substrate the shift of the Fermi level must be at least $\approx$ 40 meV.\cite{Mak2012c} The full width half maximum (FWHM) of the \textit{A}$^-$ peak is unaffected by the substrate; we find for all spectra a value of $\approx$ 100 meV. For the \textit{A} peak in the freestanding area we find a value of $\approx$ 47 meV.
The vast majority of reported photoluminescence data from supported single-layer MoS$_2$ is in the energy range of ($\approx 1.82$ eV).\cite{Splendiani2010,Plechinger2011,Bussmann2012,Conley2013a,Scheuschner2012a} 
Our results indicate that in these cases the MoS$_2$ was \textit{n}-type doped and the observed photoluminescence was originating from the \textit{A}$^-$ peak; the pure exciton \textit{A} peak is only observed in freestanding or otherwise undoped single-layer MoS$_2$.

\section{Conclusion}

We have presented a comparison of the photoluminescence spectra of suspended and supported (on Si/SiO$_2$) single- and few-layer MoS$_2$.
Freestanding single-layer MoS$_2$ shows strong photoluminescence from the \textit{A} peak (exciton) and the $A^-$ peak (trion). For single-layer MoS$_2$ on Si/SiO$_2$, the emission from the exciton (\textit{A} peak) is suppressed due to \textit{n}-type doping by the substrate; instead the photoluminescence spectrum shows only the $A^-$ peak (trion). We therefore conclude that in most cases where the reported photoluminescence of single-layer MoS$_2$ is close to the  $A^-$ energy of $\approx$1.82 eV, the MoS$_2$ is \textit{n}-type doped  and shows the \textit{A}$^-$ peak instead of the \textit{A} exciton emission. 
In few-layer MoS$_2$, the van der Waals interaction with the substrate is decreasing the \textit{I} peak energy (indirect band gap) compared to freestanding MoS$_2$. The influence of the substrate decreases strongly with increasing layer number. For bilayer MoS$_2$ we found a redshift of the \textit{I} peak position by up to $\approx$80 meV; for three to five layers it is less than 10 meV. 

Note: During revision of the manuscript, two related studies have become available.\cite{Buscema2013a,Sercombe2013}

\section{Acknowledgements}
We thank the Fraunhofer IZM (HDI \& WLP) for the preparation of the substrates. This work was supported by the European Research Council ERC under grant no. 259286 and by the SPP 1459 Graphene of the DFG. 

\bibliographystyle{apsrev4-1}

\end{document}